\documentclass[twoside,superscriptaddress,twocolumn,floatfix,a4paper,pra]{revtex4-2}

\usepackage{xcolor}
\usepackage{graphicx}
\usepackage[utf8x]{inputenc}
\usepackage[T1]{fontenc}
\usepackage{siunitx}
\usepackage{amstext}
\usepackage{hyperref}
\usepackage{amsfonts}
\usepackage{amsmath}
\usepackage{tabularx}
\usepackage{filecontents}
\usepackage{soul}
\usepackage{dsfont}
\usepackage{mathtools}
\usepackage{makecell}
\usepackage{fourier, heuristica}
\usepackage{multirow}
\usepackage{bbold}
\usepackage{fdsymbol}
\usepackage{wasysym}
\usepackage{float}

\def\block(#1,#2)#3{\multicolumn{#2}{c}{\multirow{#1}{*}{$ #3 $}}}

\begin{document}

\title {Entanglement quantification from collective measurements processed by machine learning}

\author{Jan Roik}
\email{jan.roik@upol.cz}
\affiliation{Joint Laboratory of Optics of Palacký University and Institute of Physics of Czech Academy of Sciences, 17. listopadu 12, 771 46 Olomouc, Czech Republic}

\author{Karol Bartkiewicz} \email{karol.bartkiewicz@upol.cz}
\affiliation{Institute of Spintronics and Quantum Information, Adam Mickiewicz University,
PL-61-614 Pozna\'n, Poland}
\affiliation{Joint Laboratory of Optics of Palacký University and Institute of Physics of Czech Academy of Sciences, 17. listopadu 12, 771 46 Olomouc, Czech Republic}

\author{Antonín Černoch} \email{antonin.cernoch@upol.cz}
\affiliation{Joint Laboratory of Optics of Palacký University and Institute of Physics of Czech Academy of Sciences, 17. listopadu 12, 771 46 Olomouc, Czech Republic}
   
\author{Karel Lemr}
\email{k.lemr@upol.cz}
\affiliation{Joint Laboratory of Optics of Palacký University and Institute of Physics of Czech Academy of Sciences, 17. listopadu 12, 771 46 Olomouc, Czech Republic}

\begin{abstract}
In this paper, we investigate how to reduce the number of measurement configurations needed for sufficiently precise entanglement quantification. Instead of analytical formulae, we employ artificial neural networks to predict the amount of entanglement in a quantum state based on results of collective measurements (simultaneous measurements on multiple instances of the investigated state). This approach allows us to explore the precision of entanglement quantification as a function of measurement configurations. For the purpose of our research, we consider general two-qubit states and their negativity as entanglement quantifier. We outline the benefits of this approach in future quantum communication networks.
\end{abstract}

\date{\today}

\maketitle
\section{Introduction} 

Quantum entanglement shows immense potential as a resource in various fields of research such as quantum computing \cite{Steane_1998}, quantum cryptography \cite{RevModPhys.74.145}, and quantum teleportation experiments \cite{Bouwmeester1997}. Even though entanglement has been studied for about a century now \cite{schrodinger_1935,PhysRev.47.777}, finding a method for its experimentally feasible quantification for general quantum states is still an open and hard problem \cite{Hiesmayr2021,Gurvits2003,Gharibian2010,Huang2014}.

The most robust procedure so far seams to be the full quantum state tomography \cite{Bartkiewicz2016,PhysRevA.70.052321}, subsequent reconstruction of the density matrix \cite{paris2004quantum}, and calculation of entanglement measures. These measures include negativity \cite{PhysRevA.58.883}, concurrence \cite{bennett1996mixed,hill1997entanglement} or relative entropy of entanglement \cite{vedral1997quantifying,vedral1998entanglement}. For a review see (\cite{guhne2009entanglement}). The problem of full state tomography lies in the unfavorable scaling of the number of measurement configurations as function of Hilbert space dimension. Even for a two-qubit system, one needs to apply at least 15 measurement settings while also inevitably obtaining some information on the investigated system that is irrelevant to entanglement quantification. In order to lower the number of measurement configurations, entanglement witnesses have been proposed \cite{PhysRevLett.23.880,PhysRevA.88.052105,PhysRevA.68.052101,PhysRevLett.95.240407,PhysRevLett.93.230501,PhysRevLett.109.200503,PhysRevLett.97.050501,Walborn2006,PhysRevLett.100.140403,PhysRevLett.104.210501,PhysRevLett.98.110502,Eisert_2007,PhysRevA.77.030301,PhysRevA.81.022307,PhysRevLett.107.150502,PhysRevLett.106.190502,PhysRevA.86.062329,Rudnicki_2014,PhysRevA.90.024301,PhysRevLett.105.230404,PhysRevLett.108.240501}. These instruments are, however, designed to merely detect entanglement and can be used as measures only in limited cases such as quasi-pure states. To alleviate the problem of state dependency of entanglement detection, the concept of nonlinear entanglement witnesses has been introduced \cite{PhysRevLett.96.170502}. A noteworthy class of nonlinear witnesses are the so-called collective witnesses based on simultaneous measurement on multiple instances of the investigated state \cite{PhysRevLett.107.150502,PhysRevA.86.062329}. Entanglement measures can be estimated from collective measures as well. Analysis reveals that 4 copies of a two-qubit system need to be investigated simultaneously which can prove experimentally too demanding \cite{PhysRevA.91.022323}. To overcome this challenge, we limit ourselves to having simultaneously only two copies of the investigated state. In this configuration the relation between the outcomes of a collective measurement and an entanglement measure, say the negativity, is far from trivial.

\begin{figure}
		\begin{center}
		\includegraphics[scale=0.38]{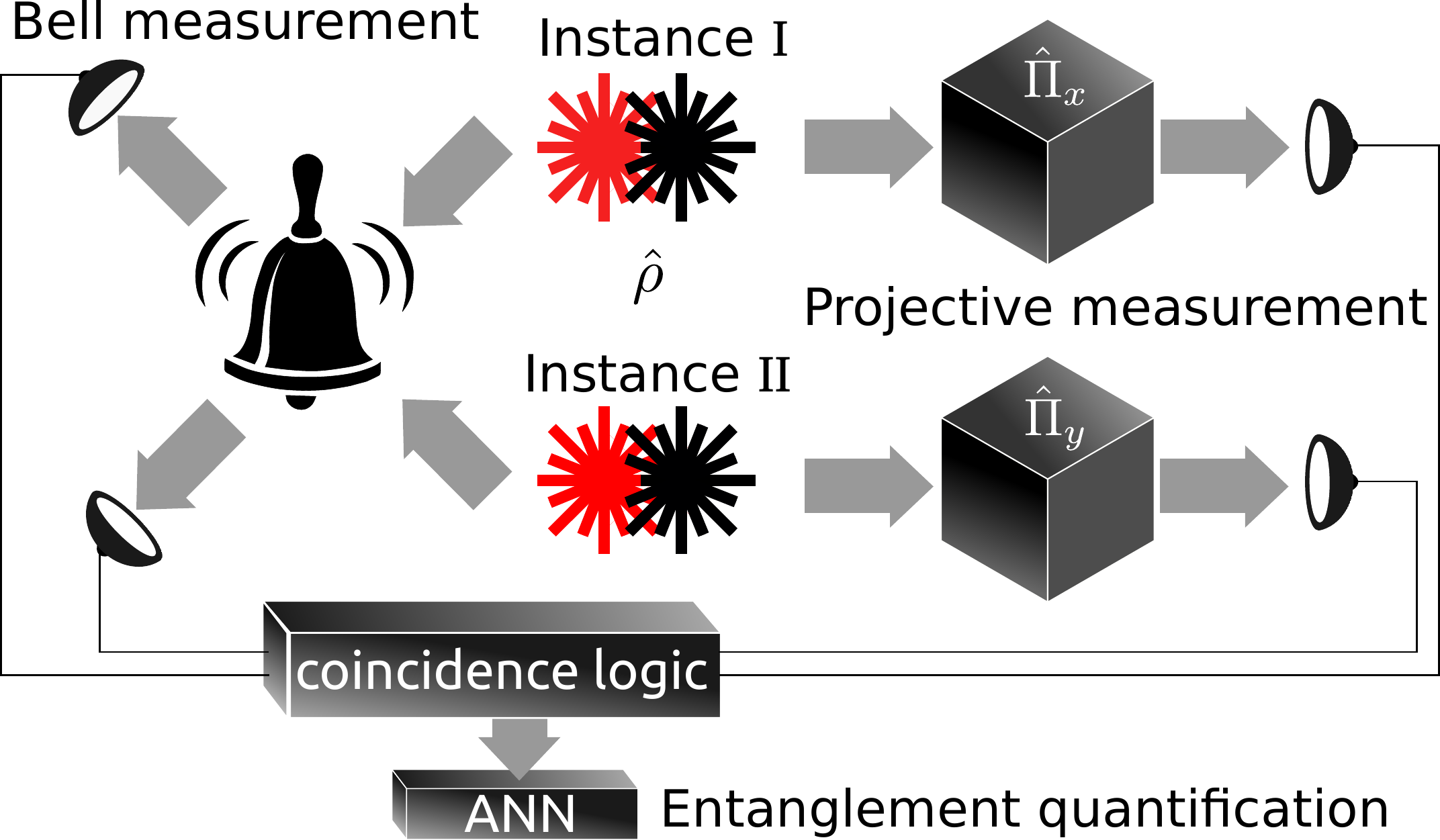}
		\caption{Scheme of a collective measurement: two instances of the investigated states $\hat{\rho}$ are subject to simultaneous measurement. While one qubit of each instance undergoes local projections, the other two qubits are nonlocally projected onto a Bell state. As explained in the text, these measurements are fed to an artificial neural network (ANN) that predicts the negativity of $\hat{\rho} $.}
		\label{fig:measurement}		
		\end{center}
\end{figure}

Machine learning has penetrated into many areas of science helping with finding complex models based on large data sets \cite{mohri2018foundations}. Artificial neural networks (ANNs) are particularly well suited for recovery of nonlinear dependencies so they have been already used to investigate properties of quantum states. Artificial intelligence was applied to entanglement detection \cite{PhysRevLett.120.240501,Ma2018,PhysRevA.98.012315}, quantification of various properties of quantum states \cite{PhysRevA.100.022314,zhang_yang_he_chen_2020,PhysRevLett.122.200401,PhysRevA.103.022425} or compressed sensing. In this paper we make use of the predictive power of an ANN to estimate quantum state negativity based on the outcomes of collective measurement.
\section{The concept of collective measurements}

Although our idea can be generalized, we focus our investigation on the entanglement of two-qubit states $\hat{\rho}$.  In order to perform collective measurements on these states, one needs to start with the preparation of two instances of $\hat{\rho}$ resulting in an overall density matrix of the entire system $\hat{\rho_4}=\hat{\rho} \otimes \text{SWAP}\; \hat{\rho}\; \text{SWAP}^{\dagger}$; where the SWAP operator interchanges the order of subsystems. [see Eq.(\ref{eq:SWAP}) in Appendix]. One qubit from each instance is projected locally, while the remaining qubits undertake a nonlocal Bell-state projection. For the visualization of this procedure, see Fig. \ref{fig:measurement}. For a given pair of local projections, the result of collective measurement is the probability of a successful singlet Bell-state projection imposed to the nonlocally projected qubits 
\begin{equation}
P_{xy}= \frac{\text{Tr}[\left(\hat{\rho}_4 \right) \left( \hat{\Pi}_x \otimes \hat{\Pi}_{\text{Bell}} \otimes \hat{\Pi}_y \right)]}{\text{Tr}[\left( \hat{\rho}_4 \right) \left( \hat{\Pi}_x \otimes \hat{\mathbb{1}}^{(4)} \otimes \hat{\Pi}_y \right)]}.
\label{eq:probability}
\end{equation}
In this equation $\hat{\Pi}_x$ and $\hat{\Pi}_y$ are local projections onto single-qubit states $|x\rangle$ and $|y\rangle$, $\hat{\Pi}_{\text{Bell}}$ denotes projection onto the singlet Bell state and $\hat{\mathbb{1}}^{(4)}$ represents four-dimensional identity matrix. One collective measurement configuration corresponds to the choice of one $\hat{\Pi}_x$ and one $\hat{\Pi}_y$.

\begin{figure}
		\begin{center}
		\includegraphics[scale=0.4]{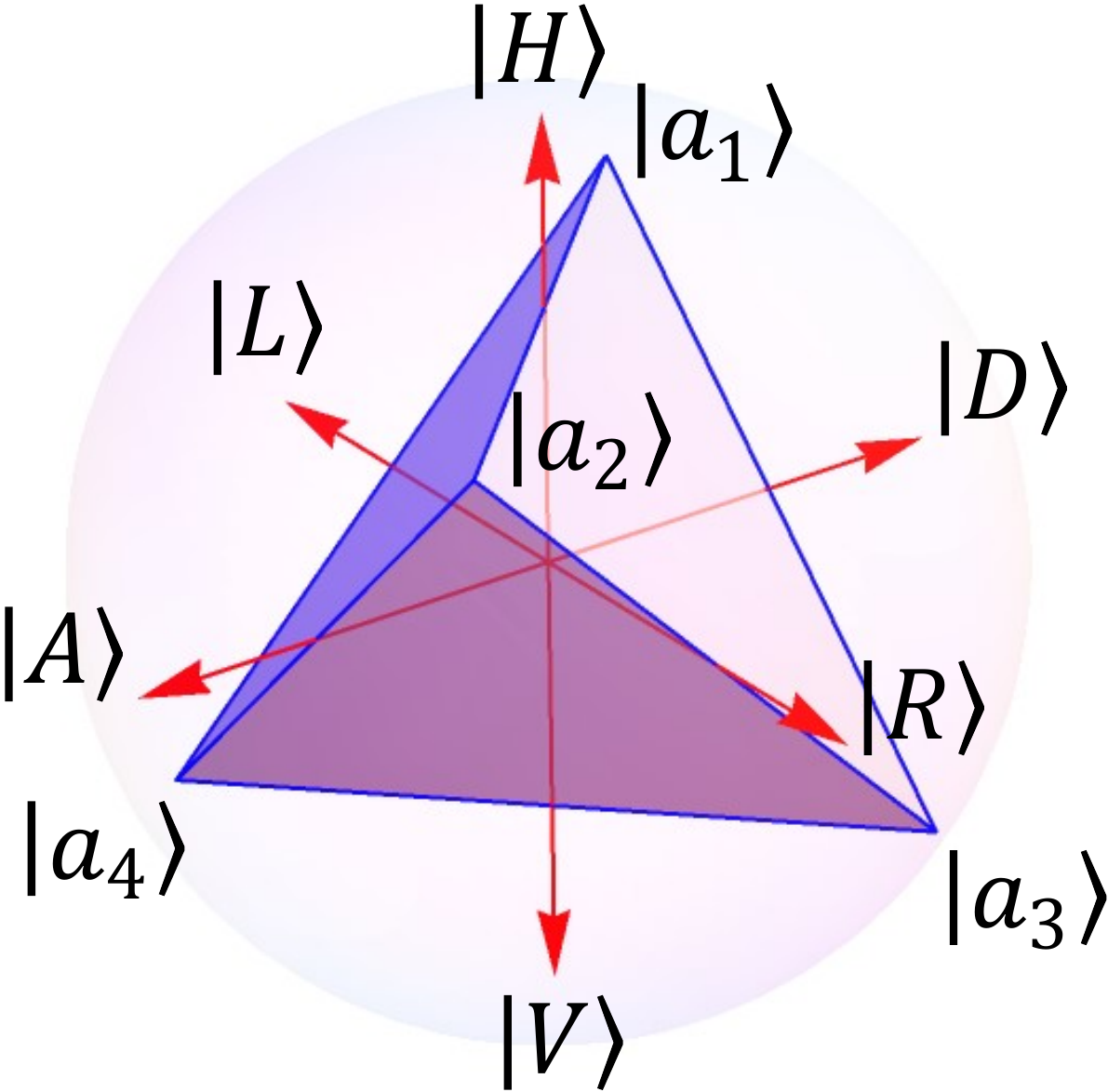}
		\caption{Minimal set of tomographic projections visualized on the Poincaré sphere by $|a_j\rangle$ vertices, where  $|a_1\rangle=\frac{1}{\sqrt{3}}(1,1,1),\:|a_2\rangle = \frac{1}{\sqrt{3}}(1,-1,-1),\:|a_3\rangle = \frac{1}{\sqrt{3}}(-1,1,-1),\:|a_4\rangle = \frac{1}{\sqrt{3}}(-1,-1,1)$. Red arrows represent $|H\rangle -\text{horizovtal},|V\rangle -\text{vertical},|D\rangle -\text{diagonal},|A\rangle -\text{antidiagonal},|R\rangle -\text{circular right-hand},|L\rangle -\text{circular left-hand}$ basis states.}
		\label{fig:tetrahedron}		
		\end{center}
\end{figure}

This paper aims at efficient entanglement quantification in two-qubit states using as few projections as possible. To achieve this goal, we take inspiration from the concept introduced by Řeháček \textit{et al.} called minimal qubit tomography \cite{PhysRevA.70.052321}. The authors established that the minimal set of tomographic projections per one qubit consists of four projections corresponding to states forming a tetrahedral inscribed into a Bloch sphere (see Fig.\ref{fig:tetrahedron}). One possible set of these projections
\begin{equation}
\begin{split}
\hat{\Pi}_1 &= \frac{1}{4}\left(\sigma_0+ \frac{1}{\sqrt{3}} \left( \sigma_x+ \sigma_y +\sigma_z\right)\right) ,\\
\hat{\Pi}_2 &= \frac{1}{4}\left( \sigma_0+ \frac{1}{\sqrt{3}} \left( \sigma_x- \sigma_y -\sigma_z\right)\right) ,\\
\hat{\Pi}_3 &= \frac{1}{4}\left( \sigma_0+ \frac{1}{\sqrt{3}} \left( -\sigma_x+ \sigma_y -\sigma_z\right)\right) ,\\
\hat{\Pi}_4 &= \frac{1}{4}\left( \sigma_0+ \frac{1}{\sqrt{3}} \left( -\sigma_x- \sigma_y +\sigma_z\right)\right) ,
\end{split}
\end{equation}
is conveniently expressed in terms of Pauli matrices 
\begin{equation}
\sigma_0=
\begin{pmatrix}
1 &0  \\ 
0 &1   
\end{pmatrix},
\sigma_x=
\begin{pmatrix}
0 &1  \\ 
1 &0   
\end{pmatrix},
\sigma_y=
\begin{pmatrix}
0 &-i  \\ 
i &0   
\end{pmatrix},
\sigma_z=
\begin{pmatrix}
1 &0  \\ 
0 &-1   
\end{pmatrix}.
\end{equation}
Using such an optimal basis, full two-qubit state tomography requires at least $15$ measurements assuming a known constant state generation rate. A density matrix $\hat{\rho}$ can be estimated from the tomography and the entanglement quantifier negativity is calculated as 
\begin{equation}
N_A= 2|\text{min}(\lambda_i)|,
\end{equation}
where $\text{min}(\lambda_i)$ is the smallest eigenvalues of partially transposed density matrix $\hat{\rho}^{PT}$ \cite{PhysRevA.58.883}.

For the collective measurement approach to be beneficial, it needs to require at most $7$ measurement configurations which is less than one-half of the projections needed for a full state tomography (in case we are using two instances simultaneously). Because of the symmetry of $\hat{\rho}_4$, the collective measurement is independent of the swap of local projections, i.e. $P_{xy}=P_{yx}$. Using this fact and considering the minimal basis set $\hat{\Pi}_{1,..,4}$, the maximal independent number of collective measurement configurations is $10$ (see Tab. \ref{tab:B}). Finding an approximated analytical formulae for quantum states negativity based on a specific number of specific collective measurement configurations is a tedious and considerably difficult task. To solve this problem, we turn to the predictive power of artificial neural networks.

\section{Artificial neural networks}
\begin{figure}
		\begin{center}
		\includegraphics[scale=0.95]{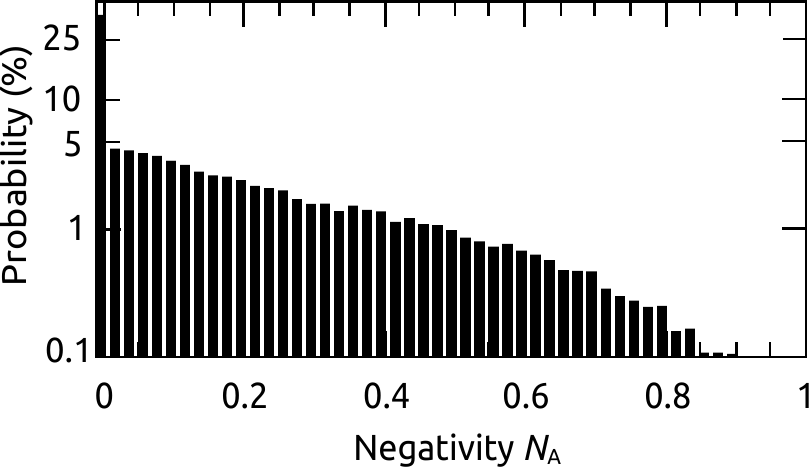}
		\caption{Distribution of negativity $N$ as the function of probability of occurrence in generated states $\hat{\rho}$.}
		\label{fig:distribution}		
		\end{center}
\end{figure}

We used TensorFlow 2.0 \cite{abadi2016tensorflow} to program the artificial neural networks (ANN) capable of quantifying the degree of the entanglement for general two-qubit states utilizing the technique of supervised learning. Various probabilities $P_{xy}$ are packed into the feature vector while negativity squared $N^2$ is used as the label. As it turns out, ANN learns more efficiently when provided with $N^2$ as a label instead of just $N$. The structure of the ANN contains two hidden layers consisting of 2000 and 1500 neurons, respectively, and uses Rectified Linear Unit (ReLU) as an activation function
\begin{equation}
\text{ReLU}(x) = x^+ = \text{max}(0,x),
\end{equation}
where $x$ is the input to a neuron. The results are processed in the final layer based on SoftPlus activation, a smooth approximation of the ReLU function
\begin{equation}
\text{SoftPlus}(x) = \ln(1+e^{x}).
\end{equation}
We used adaptive moment estimation as an optimizer and mean squared error as a loss function
\begin{equation}
\text{MSE} = \frac{1}{n} \sum_{i=1}^{n} (N_A - N_p)^2,
\end{equation}
where $n$ represents the number of all training states, $N_A$ corresponds to analytical values of negativity obtained from density matrix, and $N_p$ stands for the predicted value of negativity. We identify 100 as an optimal number of epochs for our ANN. The ANN was taught on $4\cdot10^6$ randomly generated two-qubit states $\hat{\rho}$ and tested on other unique $1\cdot10^6$ random states. For more details on random $\hat{\rho}$ generation, see Appendix. We plotted the distribution of negativity in Fig. \ref{fig:distribution}. Our ANN struck a balance between complexity and efficiency in this setting, allowing us to obtain the best results. We tested the capability of the ANN for various numbers of projections configurations $B$ from 5 to 10, i. e. the length of the feature vector is $B$. For details on exact measurement configurations used in the case of given feature dimensions $B$, see Tab~\ref{tab:B}.
\begin{table}[]
\begin{tabular}{lclll}
\hline
$B$  & \multicolumn{4}{c}{$\text{Specific projections}$}                                                                                                                   \\ \hline
5  & \multicolumn{4}{c}{$\hat{\Pi}_1 \otimes \hat{\Pi}_1,\hat{\Pi}_2 \otimes \hat{\Pi}_2,\hat{\Pi}_3 \otimes \hat{\Pi}_3,\hat{\Pi}_4 \otimes \hat{\Pi}_4,\hat{\Pi}_1 \otimes \hat{\Pi}_3$}                                                           \\
6  & \multicolumn{4}{c}{$B=5,\;\land\; \hat{\Pi}_2 \otimes \hat{\Pi}_4$}                    \\
7  & \multicolumn{4}{c}{$B=6,\;\land\; \hat{\Pi}_1 \otimes \hat{\Pi}_4$} \\
8  & \multicolumn{4}{c}{$B=7,\;\land\; \hat{\Pi}_1 \otimes \hat{\Pi}_2$} \\
9  & \multicolumn{4}{c}{$B=8,\;\land\; \hat{\Pi}_2 \otimes \hat{\Pi}_3$}                                                          \\
10 & \multicolumn{4}{c}{$B=9,\;\land\; \hat{\Pi}_3 \otimes \hat{\Pi}_4$} \\   \hline                                  
\end{tabular}
\caption{\label{tab:B}List of specific projections settings used for the learning of the artificial neural network.}
\end{table}

We intend to study the capability of ANN to quantify entanglement as the function of a number of configurations B. As mentioned above, the maximal independent number of collective measurement configurations is $10$. Therefore, we chose this case as our starting and reference point. From there, we gradually reduced the number of provided projections down to $B = 5$. The most impactful results are obtained for $B=7$ because, at that point, the number of projections drops below one-half of the projections needed for a full state tomography making this setting our primary success indicator. We used the coefficient of determination $R^2$ 
\begin{equation}
R^2=1-\frac{SS_{res}}{SS_{tot}},
\end{equation}
and standard deviation
\begin{equation}
\tau = \frac{1}{\sqrt{n}}\sqrt{\sum_{i=1}^{n}((N_A-N_p)-\mu)^2},
\end{equation}
to quantify the capabilities of the ANNs. Where the total sum of squares $SS_{tot}$ and residual sum of squares $SS_{res}$ are defined as
\begin{equation}
\begin{split}
SS_{tot}&=\sum_{i}(N_A-\bar{N})^2, \\
SS_{res}&=\sum_{i}(N_A-N_p)^2,\\
\bar{N}&=\frac{1}{n}\sum_{i=1}^{n}N_A,\\
\end{split}
\end{equation}
$\bar{N}$ represents the mean value of analytically calculated negativity, and the mean average is obtained as 
\begin{equation}
\mu = \frac{1}{n}\sum_{i=1}^{n} (N_A-N_p).
\end{equation}
   
\begin{figure*}[ht]
		\begin{center}
		\includegraphics[scale=0.85]{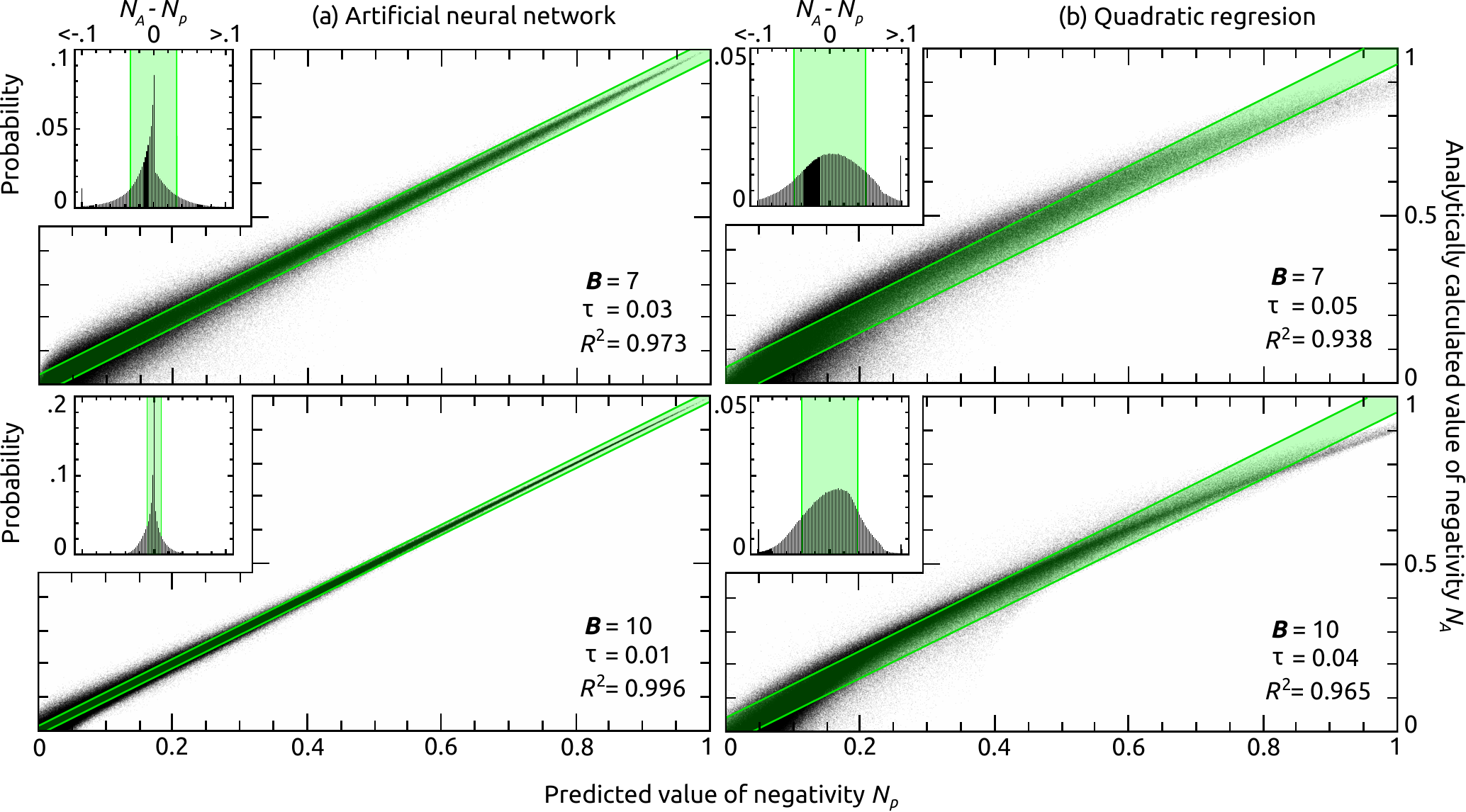}
		\caption{Comparison between analytically calculated negativity $N_A$ and negativity predicted $N_p$ by (a) artificial neural network and (b) quadratic regression for $B = 10$ and $7$ configurations respectively. In addition, the graphs include insets depicting histograms of the difference between $N_A$ and $N_p$. Every tiny black dot corresponds to one of $1\cdot10^6$ tested random states. The coefficient of determination $R^2$ and standard deviation $\tau$ are also included in the legend. In an ideal case, all dost should lie on a diagonal line $N_A = N_p$. Green stripes depict standard deviation $\pm\tau$ from such an ideal case.}
		\label{fig:ANN_VS_REG}		
		\end{center}
\end{figure*}

\section{Results.} 
\begin{figure}
		\begin{center}
		\includegraphics[scale=0.95]{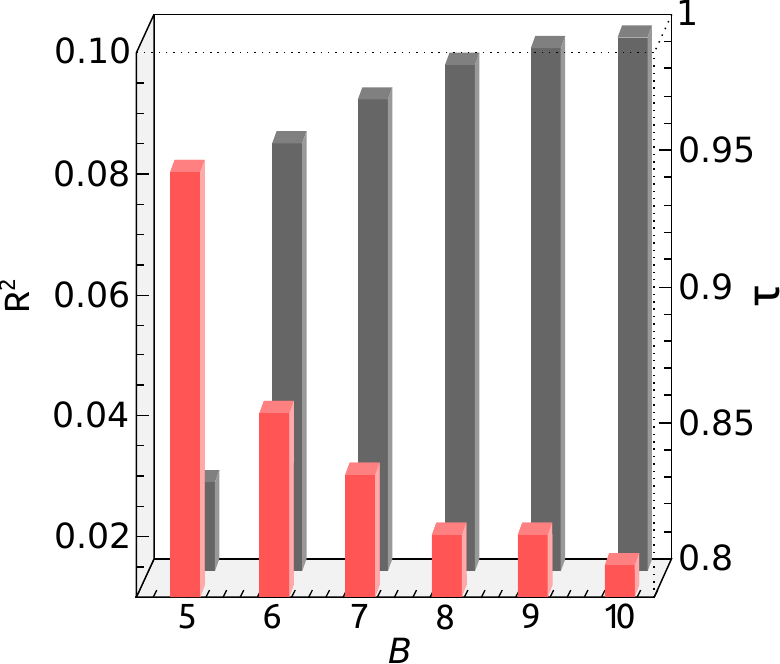}
		\caption{The coefficient of determination $R^2$ (represented by black back columns) and standard deviation $\tau$ (represented by red front columns) are plotted for all measurement configurations $B$ to visualize trends in the results.}
		\label{fig:trend}		
		\end{center}
\end{figure}

First, we provided the ANN with all available information about the investigated state (i.e., $B=10$ projections) to set the benchmark.  In this specific case, the ANN was able to reach $R^2 = 0.996$ and $\tau=0.01 $ (see Fig. \ref{fig:ANN_VS_REG}). Further evolution of the network by using a more complex structure of ANN and enlargement of the training data did not improve the performance of the ANN. We also tried to enlarge the number of epochs, but even additional epochs did not bring better results. Therefore, we conclude that we have found the best approximation of the negativity function $N(\hat{\rho})$ using only two copies of the investigated state and collective measurements. In the next step, we reduced the number of projections to $B=9$. As expected, the performance of the network decreased to $R^2 = 0.992$ and $\tau=0.02 $. Further decrease in the number of projections to $B=8$  did not reveal anything noteworthy but merely confirmed the trend established above. The performance of the ANN taught on $B=7$ projections represents the most notable result $R^2 = 0.973$ and $\tau=0.03$ (see Fig. \ref{fig:ANN_VS_REG}) because, at this point, we reduced the number of projections under the full tomography requirements. Obtained results are similar to the limits of the analytical calculations performed on estimated density matrix from actual experimental tomography data since those calculations cannot be completely accurate due to unavoidable measurement uncertainties, which usually contribute to final analytical errors by a similar margin, i.e., $\tau=0.03$ \cite{jirakova2021experimental}. When the number of projections dropped to $B=6$ we noticed some decline in the prediction capabilities ($R^2 = 0.95$ and $\tau = 0.04$). Even for $B=6$ measurements configurations, the observed prediction error is still quite comparable to experimental full state tomography. We tried to limit the number of projections as much as possible, but we drew the line at $B=5$. In this case, the ANN performance pecked at $R^2=0.83$ and $\tau =0.08$. At this point, the prediction error is already significant, and therefore we did not proceed with further decreasing of $B$. For an overview of the results, see Fig. \ref{fig:trend}.

In Fig.\ref{fig:ANN_VS_REG} we have also compared the ANN models to  quadratic regression models for $B=7,8.$ The ANNs use significantly more model parameters than our regression models, but they perform much better. The coefficient of determination for the ANN models is typically larger by $0.03$ if compared with the regression models. The typical root mean square difference between the predicted values $N_p$ of the ANNs and quadratic regression models is circa $0.17,$ and it does not depend strongly on the number of measurement configurations $B$.
However, there is also a benefit of using the quadratic regression models. By doing so we are able to directly obtain compact approximate formulas for negativity as functions of assorted collective measurements(see Appendix C).

\section{Conclusions}
The above-presented results demonstrate a significant potential of ANNs together with collective measurements for entanglement quantification. Even for $B=6$ \text{and} $7$ measurement configurations, the collective measurement performs similarly to experimental full quantum state tomography committing the predictive error of about $3\%$ (in terms of standard deviation). Considering that the particular geometry of collective measurement also overlaps with entanglement swapping setup \cite{PhysRevLett.80.3891}, implementing entanglement quantification using this configuration can prove interesting for future quantum communication networks \cite{PhysRevApplied.14.064071}. The method presented in this paper can be used for effective entanglement quantification in entanglement swapping-based communication networks. It should also be emphasized that one can, in principle, train the ANN on numerically generated quantum states and then apply such ANN on real experimental data. Such experimental investigation is, however, beyond the scope of this paper.

\begin{figure*}[ht]
		\begin{center}
		\includegraphics[scale=0.85]{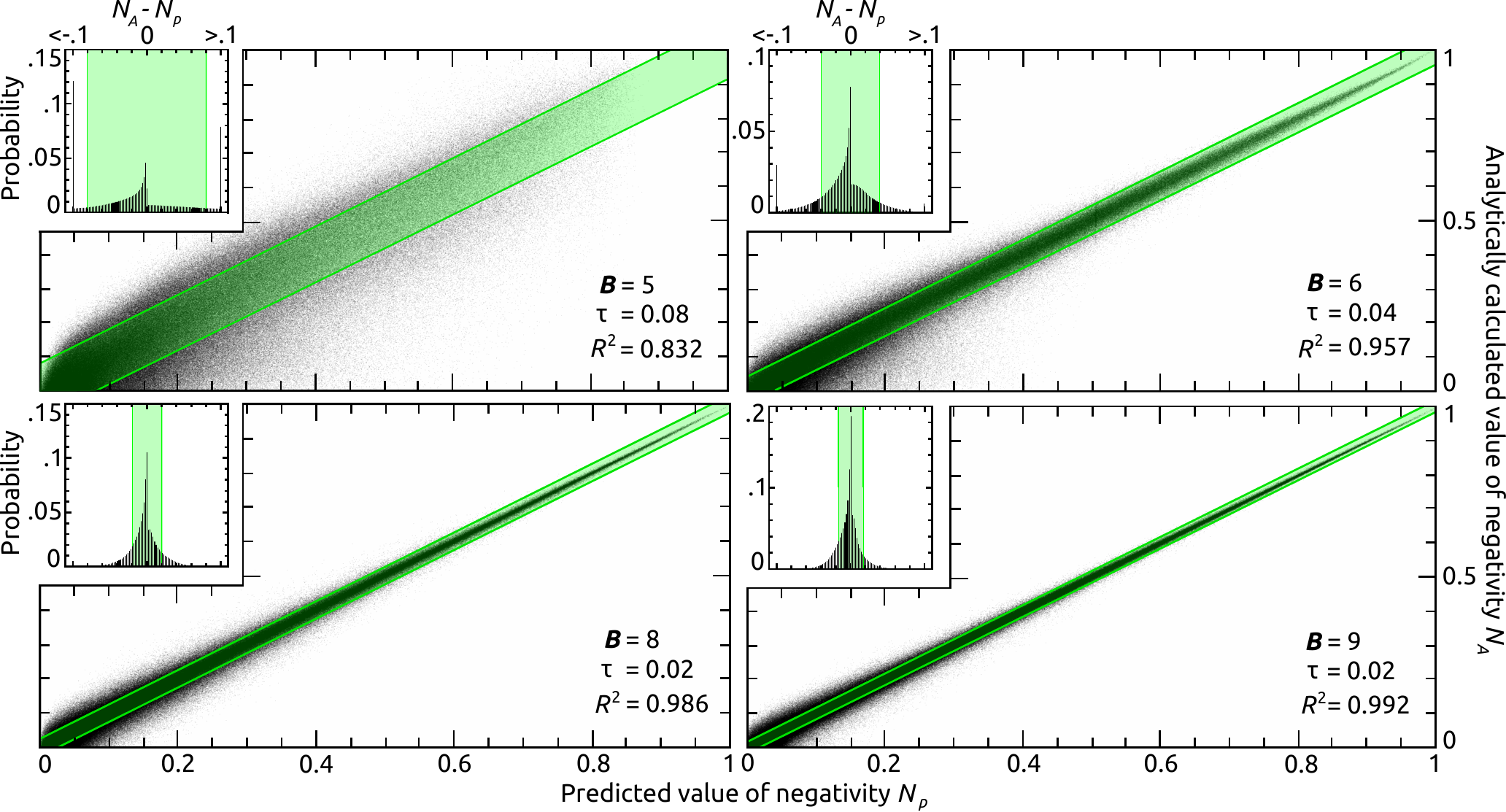}
		\caption{Comparison between analytically calculated negativity $N_A$ and negativity predicted $N_p$ by (a) artificial neural network and (b) quadratic regression for $B = 9,8,6$ and $5$ configurations respectively. In addition, the graphs include insets depicting histograms of the difference between $N_A$ and $N_p$. Every tiny black dot corresponds to one of $1\cdot10^6$ tested random states. The coefficient of determination $R^2$ and standard deviation $\tau$ are also included in the legend. In an ideal case, all dost should lie on a diagonal line $N_A = N_p$. Green stripes depict standard deviation $\pm\tau$ from such an ideal case.}
		\label{fig:N_9-5}		
		\end{center}
\end{figure*}

\section*{Acknowledgment} Authors thank Cesnet for providing data management services. Authors acknowledge financial support by the Czech Science Foundation under the project No. 20-17765S. KB also acknowledges the financial support of the Polish National Science Center under grant No. DEC-2019/34/A/ST2/00081. JR also acknowledges internal Palacky University grant IGA-PrF-2021-004. The authors also acknowledge the project No. CZ.02.1.01./0.0/0.0/16\textunderscore 019/0000754 of the Ministry of Education, Youth and Sports of the Czech Republic. Source codes, as well as trained ANN's parameters, are accessible via the digital supplement \cite{digital}.

\section*{Appendix}

\subsection{Preparation of general two-qubit states}
 Random two-qubit states were generated from $4\times 4$ diagonal matrix $\hat{\rho}_i$ according to Ref.\cite{Maziero2015} 
\begin{equation}
\label{eq:M}
\hat{\rho}_i=
\begin{pmatrix}
\rho_{11} &0  &0  &0 \\ 
0 &\rho_{22}  &0  & 0\\ 
0 &0  &\rho_{33}  & 0\\ 
0 &0  &0  & \rho_{44}
\end{pmatrix}
\end{equation}
where $\rho_{11} = r_1$; $\rho_{22} = r_2(1-\rho_{11})$; $\rho_{33} = r_3(1-\rho_{11}-\rho_{22})$; $\rho_{44}=r_4(1-\rho_{11}-\rho_{22}-\rho_{33})$; $r_n$ for $n=1,2,3,4$ are uniformly distributed random numbers from range $[0,1]$. In the next step, the proper random unitary transformation was used in order to create a density matrix of general random 2-qubit state \cite{S0219749913500159} 
\begin{equation}
\label{eq:M}
\begin{split}
U=&
\begin{pmatrix}
1 &0  &0  &0 \\ 
0 &1  &0  & 0\\ 
0 &0  &\block(2,2){U_1}\\ 
0 &0  &  & 
\end{pmatrix}
\begin{pmatrix}
1 &0  &0  &0 \\ 
0 &\block(2,2){U_2} &0\\
0 &  &  & 0\\ 
0 &0  &0  &1 
\end{pmatrix}
\begin{pmatrix}
\block(2,2){U_3} &0 &0\\
 &  &0  &0 \\ 
0 &0  &1  & 0\\ 
0 &0  &0  &1 
\end{pmatrix}\\
&\begin{pmatrix}
1 &0  &0  &0 \\ 
0 &1  &0  & 0\\ 
0 &0  &\block(2,2){U_4}\\ 
0 &0  &  & 
\end{pmatrix}
\begin{pmatrix}
1 &0  &0  &0 \\ 
0 &\block(2,2){U_5} &0\\
0 &  &  & 0\\ 
0 &0  &0  &1 
\end{pmatrix}
\begin{pmatrix}
1 &0  &0  &0 \\ 
0 &1  &0  & 0\\ 
0 &0  &\block(2,2){U_6}\\ 
0 &0  &  &
\end{pmatrix},
\end{split}
\end{equation}
where
\begin{equation}
U_j= e^{i\alpha_j}
\begin{pmatrix}
e^{i\psi_j}\cos{\phi_j}  & e^{i\chi_j}\sin{\phi_j} \\ 
-e^{-i\chi_j}\sin{\phi_j}  & e^{-i\psi_j}\cos{\phi_j}\\ 
\end{pmatrix}\\,
\;\;\;\;\;\;\;\;\; j=1,\dots,6
\end{equation}
with $0 \leq  \phi \leq  \frac{\pi}{2}$,$0 \leq \alpha, \psi, \chi < 2 \pi$.
The homogenous distribution of states was ensured by $\phi_j = \arcsin \sqrt{\xi_j}, \xi_j \in [0,1]$. Parameters $\phi_j, \psi_j, \chi_j, \alpha_j$ and $\xi_j$ are picked from their respective intervals with uniform probability. The final density matrix was obtained as $\hat{\rho} = U\hat{\rho}_i U^\dagger$. To mathematically describe the collective measurement, a 4-qubit density matrix of the entire system $\hat{\rho}_4$ was defined as $\hat{\rho}_4 = \hat{\rho} \otimes SWAP\; \hat{\rho}\; SWAP$
where
\begin{equation}
SWAP = \begin{pmatrix}
1 & 0  & 0  & 0 \\ 
0 & 0  & 1  & 0 \\ 
0 & 1  & 0  & 0 \\ 
0 & 0  & 0  & 1 
\end{pmatrix}.
\label{eq:SWAP}
\end{equation}

\subsection{Results for the additional number of measurement settings $B$}
Results of all additional measurement settings $B=9,8,6$ and $5$ are depicted in Fig. \ref{fig:N_9-5}.
\subsection{Quadratic fit of the negativity function}
The quadratic regression model for the negativity can be formally written as
$N_p  = \vec{\theta}_B\cdot\vec{x},$
where the vector $\vec{x} = (1,x_1,x_2,...,x_B,
                 x_1^2,x_1x_2,...,x_1x_B,
                 x_2^2,x_2x_3,...,x_2x_B,
                 ...,
                 x_B^2)$
contains experimental results $x_1,x_2...,x_B.$ The optimized quadratic model parameters
read:

$\vec{\theta}_5=$ 
(0.0000,  -3.5815,    0.7204,  -3.7203,    0.9883,    5.0442,  -1.2165,    2.2688,    4.7236,    2.0851,    5.9580,  -0.7245,    1.5130,  -3.1161,  -5.0063,  -0.9646,   2.2562,    6.5298,  -1.2991,  -5.6469,  -8.1649),

$\vec{\theta}_6=$
(-0.0000,  -1.3150,  -1.3392,  -1.3420,  -1.3102,    2.1782,    2.1709,  -1.5659,    3.0431,  -0.8994,    3.3200,    5.2307,  -5.2516,  -1.2834,    2.9993,  -0.9053,  -5.2575,    5.0915,  -1.2830,    3.0224,    5.0876,  -5.2393,  -1.5616,  -5.2370,    5.2126,  -3.8903,    4.6612,  -3.8750),

$\vec{\theta}_7=$
(0.0000,  -1.8673,  -1.0936,  -1.0942,  -1.8643,    1.9973,    1.9907,    0.6725,  -2.3285,    3.2386,    0.1532,    3.3836,    2.9791,  -5.8464,    4.5824,  -0.9105,    1.0969,    0.1474,  -3.7233,    5.0767,  -2.1132,  -0.9173,    3.2317,    5.0804,  -3.7180,  -2.1183,  -2.3273,  -5.8377,    2.9738,    4.5803,  -3.9245,    3.6918,    1.6918,  -3.9061,    1.6854,  -3.2525),

$\vec{\theta}_8=$
(0.0000,  -2.2433,  -1.5432,  -0.9637,  -1.6833,    1.8790,    1.9100,    0.6344,    0.5783,  -2.4672,    3.2790,    0.1640,    3.7855,    1.8370,  -6.5895,    3.9005,    3.6195,  -1.4810,    1.0215,    1.5379,  -3.1384,    3.3574,  -3.9152,    4.0323,  -0.6074,    1.8475,    4.2688,  -2.1111,  -1.2269,  -1.1623,  -2.2688,  -3.7386,    3.8779,    4.6486,  -4.1114,  -3.1326,    1.5010,    1.1220,    0.1660,  -5.0920,    3.0011,    2.9862,  -4.0649,    2.1316,  -3.5849),

$\vec{\theta}_9=$
( 0.0000,  -1.8418,  -1.8173,  -1.5094,  -1.5388,    1.1660,    1.1446,    1.2874,    0.2835,    1.2762,  -1.9965,    3.2911,    1.1961,    1.7895,    2.3318,  -3.9072,    2.7700,    3.3719,  -3.2130,  -1.8183,    1.5827,    1.1717,  -3.8511,    2.2402,  -3.2112,    3.4184,    2.6514,  -1.2886,    1.4645,    3.2473,  -1.7011,  -0.9769,  -2.6345,    3.4001,  -1.5136,  -1.7254,    3.4717,    3.4980,  -2.8135,  -0.8313,  -3.7572,  -0.0293,    1.3468,    1.8763,    2.5188,  -3.7814,    2.5555,    2.0001,    1.2824,  -3.9921,    2.1486,  -1.7069,  -4.1664,    2.1724,  -3.9370),

$\vec{\theta}_{10}=$
(0.0000,  -1.7756,  -1.7476,  -1.7491,  -1.7733,    0.7967,    0.7907,    1.0549,    0.7877,    1.0520,    0.7919,  -1.5574,    1.3280,    1.3013,    1.9322,    2.4272,  -2.0933,    2.6149,    2.4268,  -1.2658,  -2.0948,  -1.3819,    1.8574,    1.3061,  -2.1262,    2.3244,  -1.2756,    2.3518,    2.5663,  -2.0953,  -1.3768,    1.3227,    2.3354,  -2.1160,  -1.2689,  -2.1023,    2.5775,    2.3250,  -1.5537,  -2.0870,    2.4221,    2.6095,  -2.1080,  -1.2516,    2.4193,  -4.7497,  -3.1170,    3.0234,    2.6724,    3.0255,    2.7072,  -4.7590,    3.0285,    2.7435,    3.0374,    2.6720,  -5.1603,    3.0316,  -5.4810,    3.0262,  -4.7573,    3.0290,  -3.1225,  -5.1586,    3.0336,  -4.7460).

\begin{table}[]
\begin{tabular}{ccccc}
\hline
\multicolumn{1}{l}{} & \multicolumn{2}{c}{ANN} & \multicolumn{2}{c}{REG}            \\
B                    & $R^2$      & $\tau$     & $R^2$ & \multicolumn{1}{l}{$\tau$} \\ \hline
5                    & 0.832      & 0.08       & 0.809 & 0.09                       \\
6                    & 0.957      & 0.04       & 0.926 & 0.06                       \\
7                    & 0.973      & 0.03       & 0.939 & 0.05                       \\
8                    & 0.986      & 0.02       & 0.947 & 0.05                       \\
9                    & 0.992      & 0.02       & 0.959 & 0.04                       \\
10                   & 0.996      & 0.01       & 0.966 & 0.04                       \\ \hline
\end{tabular}
\caption{\label{tab:ANN_VS_REG}Comparison of the results obtained by ANN and REG (regression) for B = 5-10. Where $R^2$ represents the coefficient of determination and $\tau$ stands for standard deviation.}
\end{table}

\newpage
\bibliographystyle{ieeetr}


\end{document}